\documentclass[12pt]{article}
\usepackage{epsfig,graphicx}

\title{Nonperturbative  approach to the parton model}
\author{  Yu.A.Simonov,\\ Institute of Theoretical and Experimental
Physics\\ 117218, Moscow, B.Cheremushkinskaya 25, Russia}

\newcommand{\be}{\begin{equation}}
\newcommand{\ee}{\end{equation}}

\def\la{\mathrel{\mathpalette\fun <}}

\def\fun#1#2{\lower3.6pt\vbox{\baselineskip0pt\lineskip.9pt
\ialign{$\mathsurround=0pt#1\hfil ##\hfil$\crcr#2\crcr\sim\crcr}}}

\newcommand{{\SD}}{\rm SD}

\newcommand{\vex}{\mbox{\boldmath${\rm x}$}}
\newcommand{\vey}{\mbox{\boldmath${\rm y}$}}
\newcommand{\ver}{\mbox{\boldmath${\rm r}$}}

\newcommand{\veP}{\mbox{\boldmath${\rm P}$}}
\newcommand{\vep}{\mbox{\boldmath${\rm p}$}}

\newcommand{\vez}{\mbox{\boldmath${\rm z}$}}

\newcommand{\vek}{\mbox{\boldmath${\rm k}$}}

\newcommand{\veu}{\mbox{\boldmath${\rm u}$}}

\newcommand{\verho}{\mbox{\boldmath${\rm \rho}$}}

\newcommand{{\Mc}}{\mathcal{M}}

\newcommand{\lan}{\langle}
\newcommand{\ran}{\rangle}

\begin{document}

\maketitle
\begin{abstract}
The nonperturbative parton distributions, obtained from the Lorentz contracted
wave functions, are analyzed   in the formalism of many-particle Fock
components and  their properties are  compared to the standard perturbative
distributions. We show that the collinear and IR divergencies specific for
perturbative evolution treatment are   absent in the nonperturbative version,
however for large momenta< $ \vep^2_i \gg \sigma$ (string tension), the
bremsstrahlung kinematics is restored. A preliminary discussion of possible
nonperturbative effects in DIS and high energy scattering is given, including
in particular a possible role of multihybrid states in  creating ridge-type
effects.

\end{abstract}

\section{Introduction}
The standard partonic model \cite{1,2, 2a, 3,4} is a basic element of the
modern QCD and is widely used in the treatment of high-energy processes, see
\cite{5,5a,5b} for a recent review.

The main principle of the standard partonic model is the assumption of the
almost free relativistic motion of quarks and gluons with only perturbative
interaction mechanism of their evolution, given by the DGLAP equations
\cite{6,6a,6b}. In this process, however, as a necessary element occur IR and
collinear singularities, which are cured by making appropriate cutoffs,
implying that perturbative laws are effective everywhere in the considered
region. The neglect of nonperturbative (np) treatment is due  to several
reasons.  First of all, the processes under consideration are in the systems,
moving with high velocity, and till recent time one could treat those only in
the perturbative manner, using Feynman diagrams for free elementary
constituents. Secondly, the study of the behavior of high-excited systems even
in the c.m. system in the np approach is not well developed, and the standard
np rules here are absent.

An approach, where perturbative and np methods to the processes with high
momentum particles were combined, namely the factorization method, is well
developed, but  needs some corrections \cite{7,7a}. Therefore there is a strong
need for the development of the np side of the  high energy QCD, using in
particular methods available for relativistic np treatment.

During last decades there had  been  serious efforts in formulating the path
integral methods for QCD  based on a general
 dynamics, including perturbative
and np contributions \cite{8,8a,8b,8c,8d,8e,8f}. Recently a most advanced
version, including also external QED fields, has  appeared \cite{9,9a}, and the
spin-dependent dynamics was treated on the same ground \cite{10}.

Basing on this powerful method, the author was able recently to consider the
QCD dynamics in the fast moving system, and has discovered, that exploiting the
well-known   Lorentz  contraction effect one immediately obtains the parton
scaled distribution from the  original np wave function in the c.m. system
\cite{11}. In this way the knowledge of the np wave function, or more
generally, of the total Fock column of the system in question allows to write
down the corresponding parton distribution functions (pdf).

Explicit examples of the unpolarized valence and sea quark and gluon  pdf have
been given in \cite{12} in comparison with those calculated before in
\cite{5,5a,5b} and the similarity of the obtained   pdf with the pdf from
\cite{5,5a,5b} is striking.

Moreover it was shown in \cite{12}, that the polarized pdf of the proton can be
obtained from the np proton wave functions calculated in the full  three-quark
Dirac bispinor formalism (the so-called factorization approach).

The important point stressed in \cite{12}, is that pdf can be obtained from the
Fock components both in the c.m. and in the fast moving Lorentz frame and the
Fock composition coefficients (relative probabilities of components) are boost
invariant.

Moreover for the polarized pdf it was suggested in \cite{12} to use the
multihybrid model for the excited DIS states, which yields $\frac12 \lan
\Sigma_3\ran = 0.182$ for $Q^2=10$ GeV$^2$ which is not far from the data
(respectively 0.26 for  $Q^2=3$ GeV$^2$).

 Thus it might be a possible way for the  solution of
 the  old proton spin problem \cite{13,13a,13b}. Moreover,  the DIS generated polarized pdf are
argued to refer not to the ground state proton, but rather to a high excited
baryon state with an indefinite spin   and large gluon admixture.

In the present paper we write down explicit connection of pdf with Fock
components and stress the  possibly important role of multihybrid states for
pdf of high excited hadrons. We henceforth study the Hamiltonian and wave
functions of hybrids and multihybrids and their contribution to pdf's. Finally
we turn to the evolution process and make the np analysis of IR and collinear
effects, which are finite in the np framework and divergent in the perturbative
formalism.  We demonstrate how the perturbative picture develops at high
momenta, $p \gg \sqrt{\sigma}$, and   how it is goes over into the $np$ regime
at low momenta.

The paper is organized as follows. In section 2 the Fock component structure of
pdf is outlined, in section 3 the hybrid and multihybrid states are analyzed
and the gluon and quark pdf are  calculated in comparison with known data. In
section 4 the IR and collinear effects are studied both in the perturbative and
np language. Section 5 contains summary and discussion.

Two appendices contain details of derivation. In appendix 1  the path-integral
calculation of the  standard IR and collinear amplitudes demonstrates the
absence of divergencies in the np dynamics. The appendix 2 contains calculation
of the triangle and quadratic Feynman diagrams with confinement, which  shows
how perturbative dynamics occurs for large momenta, $p^2\gg \sigma$.

\section{Parton distributions from Fock components}

We follow here the formalism introduced in \cite{12,14,15} and write the total
Hamiltonian and the Fock wave function as follows

\be \hat H = \left( \begin{array} {lll} H_{11},& V_{12},& ...\\ V_{21}, &
H_{22},&...\\
...&...&...\end{array}\right); ~~
 \hat H\Psi_N=(\hat H^{(0)}+\hat V) \Psi_N,\label{1}\ee
where $\Psi_N$ is decomposed in Fock components as follows \be
  \Psi_N=\sum_{m\{k\}}c^N_{m\{k\}}\psi_{m\{k\}},~~\psi_n (P, \xi, k) \equiv \psi_{n,\{k\}}\label{2}\ee

Here $m\{k\},\xi$ refer to the number and types of constitients ($q,\bar q,
g)$, modes of excitations, spin and momentum projections etc.

The orthogonality conditions look like

 \be
\int\Psi^+_N\Psi_M d\Gamma = \sum_{m\{k\}} c^N_{m\{k\}} c^M_{m\{k\}} =
\delta_{NM} .\label{3}\ee

  \be
  \int\psi^+_{m\{k\}}\psi_{n\{p\}} d\Gamma
  =\delta_{mn}\delta_{\{k\}\{p\}}.\label{4}\ee From (\ref{1}) one finds the equation for $c^N_{m\{k\}}$

  \be c^N_{n\{p\}} (E_N-E^{(0)}_{n\{p\}}) = \sum_{m\{k\}} c^N_{m\{
  k\}}V_{n\{p\}, m\{k\}}\label{5}
  \ee
 where
  \be
  V_{n\{p\}, m\{k\}}= \int \psi^+_{n\{p\}}\hat V \psi_{m\{k\}}
  d\Gamma.\label{6}\ee

 The Fock component approach was exploited earlier in \cite{13x}, see
 \cite{13xx} for more references, in the framework of the light-cone formalism
 and perturbative dynamics for the polarized parton distributions. Below we shall handle the
  np components in the  unpolarized case, and
 in the  next section specifically for the multihybrid states.

In what follows we shall have in mind the large $N_c$ limit, classifying
different contributions to $\Psi_n$, namely in the  leading order all states
are made of gluons only (glueballs) or of  hybrids: $q, \bar q$ plus any number
of gluons. In the next order of the  $ 1/N_c$  expansion the decay into two
bound states due to the process $g\to q \bar q$ is possible.

We are using the following normalization condition as in \cite{12} for the $n$
particle bound state. \be \int| \tilde \varphi_n (\vek^{(1)}, \vek^{(2)}, ...
\vek^{(n)}|^2 \prod_i \frac{d^3\vek^{(i)}}{(2\pi)^3} (2\pi)^3 \delta^{(2)}
(\sum_i \vek_\bot^{(i)}) \delta (1-\sum_i  x_i) =1 \label{7}\ee and
$dk_\|^{(i)} = M_0 dx_i$, where $M_0$ is the c.m. energy (mass) of system.
Writing (\ref{7}) as $\int|\tilde \varphi_n|^2 d \tau_n =1$, one can write the
$u,d,$ and $g$ distributions  trough the Fock components $\tilde \varphi_{3q},
\tilde \varphi_{4q\bar q}, \tilde \varphi_{3qg},$ as follows ($4q\bar q=
uuud\bar u$)
$$ u(x, k^2_\bot) = \sum_{n=3q,4q\bar q}  u_n(x, k^2_\bot) = |c_{3q}|^2
\int|\tilde \varphi_{3q} |^2 d \tau_{3q} \sum_{i=1,2} \delta^{(2)}
(\vek_\bot^{(i)}  - \vek_\bot) \delta (x_1-x)+$$ \be +|c_{4q}|^2 \int|\tilde
\varphi_{4q\bar q} |^2 d \tau_{4q\bar q}  [ \sum_{i=1}^3 \delta^{(2)}
(\vek_\bot^{(i)}  - \vek_\bot) \delta (x_i-x)+\label{8}\ee
$$+|c_{3qg}|^2 \int|\tilde
\varphi_{3qg} |^2  \sum_{i=1,2}^3 \delta^{(2)} (\vek_\bot^{(i)}  - \vek_\bot)
\delta (x_i-x)+...,$$

\be  \bar u(x, k_\bot^2)=\int|\tilde \varphi_{4q\bar q} |^2 d \tau_{4q\bar q}
 \delta^{(2)} (\vek_\bot^{(5)}  - \vek_\bot) \delta (x_5-x)+...,\label{9}\ee

 \be  g(x, k^2_\bot)== \int|\tilde
\varphi_{3qg} |^2 \delta\tau_{3qg} \delta^{(2)} (\vek_\bot^{(4)}  - \vek_\bot)
\delta (x_4-x)+... . \label{10}\ee

Assuming equal contributions of additional $q$ and $\bar q$, one can write  the
standard  relations \be u^v (x, k^2_\bot) = u(x, k^2_\bot) - \bar u
(x,k^2_\bot)\label{11}\ee \be \int u^v (x, k^2_\bot) d^2 \vek_\bot dx = 2 (
|c_{3q}|^2 + |c_{4q\bar q}|^2 + |c_{3qg}|^2+...) = 2,\label{12}\ee where we
have used the total orthonormality condition (\ref{3}), a similar relation for
the $d$ quark, $d^v (x, k^2_\bot)$ obtains from (\ref{8}), (\ref{9}), replacing
the sums $\sum_{i=1,2}$ and $\sum^3_{i=1}$ by the terms with $i=3$ and $i=4$
respectively, resulting in the normalization equation \be \int d^v (x,
k_\bot^2) d^2 \vek_\bot dx = |c_{3q}|^2 + |c_{4q\bar q}|^2 + |c_{3qg}|^2+...
.=1\label{13}\ee

From (\ref{12}), (\ref{13})  one easily obtains the Gross-Llevellin-Smith and
Adler relations for $u^v$ and $d^v$.

We now turn to the momentum relations, and to this end we make evident the $x$
dependence of each pdf, namely, as shown in \cite{12}, Eqs. (17), (18), the
dependence of $\tilde \varphi$  on $p^{(i)}_\bot, x_i$ enters in the form,
where there are present the c.m. values of the $i$-th longitudinal momentum
$p^{(0)}_{\| i}$ and the c.m. energy of the $i$-th  quark, antiquark or gluon,
namely

\be x_i = \frac{p_{\| i}}{P} = \frac{p^{(0)}_{\| i} +
v\varepsilon_i^{(0)}}{P\sqrt{1-v^2}}, ~~ P\sqrt{1-v^2} = M_0\label{14}\ee or
\be p^{(0)}_{\| i} = M_0 \left( x_i-\frac{\varepsilon_i^{(0)}}{M_0}\right), ~
v\to 1\label{15}\ee and writing in $\tilde \varphi$ only momenta of the $i$-th
particle, one has \be \tilde \varphi (\vep^{(0)}_{\bot i} , p^{(0)}_{\| i}) =
\tilde \varphi (\vep^{(0)}_{\bot i}, M_0 (x_i - \nu_i)), \label{16}\ee where
$\nu_i = \frac{\varepsilon_i^{(0)}}{M_0} $ is  a part of the total mass
contributed by the $i$-th c.m. energy. The $|\tilde \varphi|^2$ depends on
$x_i$ in the contribution $(\vep^{(0)}_{\bot i} )^2 + (M_0 (x_i - \nu_i))^2 =
\vep^2_i.$

The parton distribution in the hadron is expressed via the np boundstate wave
function $\varphi_N$ as

$$ D^q_h (x,k_\bot) =\sum_N |c^N|^2 \prod^N_{r=1} \frac{d^2\vek_{\bot r}
dx_r}{(2\pi)^3} \delta^{(2)}\left(\sum\vek_{\bot}\right)\delta\left(1-\sum^N_1
x_r\right)\times $$ \be \times M_0^N |\tilde \varphi_N|^2 \sum_j\delta^{(2)}
(\vek_\bot-\vek_{\bot j})\delta (x-x_j)\label{17}\ee and it satisfies the
normalization conditions

\be \int d^2 k_\bot dx D^q_H (x, k_\bot) = N_h^j\label{18}\ee

\be \int d^2 k_\bot x dx D^q_H (x, k_\bot) =1.\label{19}\ee

Taking into account (\ref{12}), (\ref{13}), (\ref{19}), one can write for the
proton pdf's the normalization condition \be \int^1_0 dx x [ u(x) + d(x) +\bar
u(x) + \bar d(x) + g(x)]=1.\label{20}\ee

We now turn to the np description of the Fock components $\tilde \varphi_N
(\vek_{1,...} \vek_n)$.

\section{Multihybrid Fock components of a hadron}

It is clear, that the general equation (\ref{5}) for the Fock components \be
c^N_{n\{p\}} (E_N - E^{(0)}_{n\{p\}}) = \sum_{m\{k\}} c^N_{m\{k\}} V_{n\{p\},
m\{k\}}\label{21}\ee contains different stages of evolution $n\{p\}$, each of
these is the one or more bound states of $q, \bar q$ and $g$, i.e. of mesons,
baryons, glueballs, and hybrids. In the lowest order of the expansion in
$1/N_c$, initial (primordial) mesons and baryons are connected only to
mesohybrids and baryohybrids respectively, containing  arbitrary number of
gluons. In the next order of $1/N_c$ one of the gluons can create the $q\bar q$
pair and split the total bound state into hadrons etc. Therefore it is  of
interest to study the spectrum and wave functions of multihybrids and we shall
do it for a mesonic multihybrid, see \cite{16,16a,16b,16c} for earlier
discussion of one-gluon hybrids, \cite{17,17a,17b,17c,17d,17e,17f}  for the
theory of hybrids in the field-correlator formalism,
\cite{17*,17*a,17*b,17*c,17*d,17*e,17*f,17*g,17*h,17*i} for the treatment in
the framework of potential and QCD sum rule methods, \cite{18,18a} for lattice
data, and \cite{19} for a review.

We consider as in \cite{14,15}  (Fig. 1) the system of $n$ gluons on the
string, connecting quark $(i=1)$ and antiquark $(i=n+2)$ and we are using as it
is usual for the path-integral Hamiltonian method \cite{9,9a,10}, the einbein
form, with effective masses $\omega_a, \omega_b$ for $q\bar q$ and $\omega_i,
i=1,...~n$ for  gluons and $\nu_k$-- parameters, replacing the linear
confinement pieces between $q,g, ..., \bar q, ~~\sigma|\ver_k|, ~~ k=1,...~
n+1$, by the quadratic terms $\sigma |\ver_k|\approx \frac12\left(\nu_k +
\frac{\sigma^2\ver^2_k}{\nu_k}\right),$ and $\nu_{k}$ to be found from the
stationary point analysis of the total energy -- this procedure is known  to
yield accuracy of the order $5\div 10\%$.

\noindent
\begin{center}

\begin{figure}
\begin{center}
  \includegraphics[width=10cm, ]{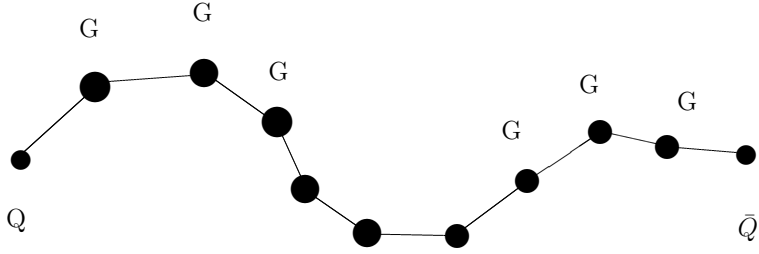}

\end{center}
  \caption{The multihybrid meson configuration with $n$ gluons inside a confining string}

\end{figure}
 \end{center}

The total Hamiltonian (without gluon exchange and spin-dependent terms) is

\be H(\veP) = \sum_{i=a,b}\frac{\vep^2_i+m^2_i}{2\omega_i} + \frac{a}{2} +
\frac{\veP^2}{2a} + \frac{\sigma^2}{2} \sum^{n+1}_{k=1} \frac{\ver^2_k}{\nu_k}
+ \sum^n_{i=1} \frac{\vep^2_i}{2\omega_i}\label{22}\ee where
$a=\sum^{n+2}_{i=1}\omega_i + \sum^{n+1}_{k=1}\nu_i$, and $ \omega_i, \nu_i$
are to be found from the stationary point conditions \be \frac{\partial
H}{\partial \omega_i } = \frac{\partial H}{\partial\nu_i} =0 \to \frac{\partial
E (\veP)}{\partial\omega_i} =  \frac{\partial E
(\veP)}{\partial\nu_i}=0.\label{23}\ee

It is easy to see, that using the l.h.s. of (\ref{23}) one arrives at the
``relativistic model of hybrids'' in the sense of Godfry and Isgur \cite{20},
we shall instead use the r.h.s. form, which will give us a simple result for
$E(\veP)$ and wave functions (the ``einbein method'').

We start with the c.m. system, $\veP=0$, and introduce a trial wave function

\be \Psi_0 = N {\exp} \left[- \sum^{n+1}_{k=1} \alpha_k \ver^2_k \right], ~~
N=\prod^{n+1}_{k=1} \left(\frac{\pi}{2\alpha_k}\right)^{-3/4}.\label{24}\ee

Leaving details of the calculation to the appendix 1, we write below the result
of the minimization  (denoted with the superscript (0)) with respect to
$\{\nu_k, \omega_i, \alpha_k\}$ in the limit $m_a =m_b=0, ~n\to\infty$ \be
\omega_i^{(0)} = \nu_i^{(0)} =\left(\frac{3\sqrt{2}}{2}\sigma
\right)^{1/2}\approx 0.62~ {\rm GeV}~\equiv \omega_0, ~~ \alpha_i^{(0)}
=\frac{\sigma}{2\sqrt{2}}\label{25}\ee \be M^{(n)}_{\min (\alpha, \nu,
\omega)}= 2\omega_0 n+O(1/n)\cong1.24~ n ~{\rm GeV}.\label{26}\ee

The average intergluon distance is found to be \be \ver_0 \equiv \sqrt{ \lan
\ver^2_k\ran} = \sqrt{\frac{3}{\sqrt{2}\sigma}}=0.68 ~{\rm fm}.\label{27}\ee

The value (\ref{26}) is an upper limit, and the actual mass should be lower. To
check the accuracy of our calculations we consider the case of the simplest
hybrid state: $ q\bar q g$, in which case the same type of the trial wave
function yields $M^{(1)}\cong 2.1$ GeV, while more accurate calculations
\cite{17} give the lowest mass $M^{(1)}\approx 1.54$ GeV.

Hence one can conclude, that the energy interval between the states of $n$ and
$(n+1)$ hybrids (one additional gluon) is

\be \Delta M= M^{(n+1)} - M^{(n)}\approx 0.8 \div 1~{\rm GeV}.\label{28}\ee

The energy intervals  of the transverse vibrational modes $\Delta E^{(TVM)}$
and longitudinal vibrational modes $\Delta E^{(LVM)}$ have been calculated in
\cite{17}  last reference, for  the case of the gluon string with fixed ends
and are equal \be \Delta E^{(LVM)} = \frac{0.53~{\rm GeV}}{[R~({\rm
fm})]^{1/3}}, ~~ \Delta E^{(TVM)} = \frac{0.69~{\rm
GeV}}{R~{\rm(fm)}}.\label{29}\ee

From (\ref{29}) one can conclude, that the lowest types of excitations for long
multigluon strings are  of the TVM and LVN types, while spin-spin interaction
interval $\Delta M_{ss}$ is of the order of 0.3 GeV. It is understandable, that
for large $n$ there appear collective excitations of the TVM and LVM types.

We now turn to the partonic form of our $n$-gluon hybrid wave function, from
(\ref{24}) one has in the $k$-space \be \tilde \varphi^2_N (\vek_1, ...
\vek_{N-1})=\frac{1}{M_0} \left( \frac{2\pi}{\alpha}\right)^{\frac32(N-1)} \exp
\left( - \sum^{N}_{i=1} \frac{\vek^2_i}{2\alpha}\right), ~~ N=n+2.\label{30}\ee

Here $\tilde \varphi_n$ is normalized as in(\ref{7}), $M_0$ is the total energy
in the c.m. system.

From (\ref{30}) one can deduce    that  the wave function for  the $i$-th
degree of freedom in (\ref{30}) can be  written in the  form of (\ref{16})
$\tilde\varphi^2 (\vek^2_i) = \tilde\varphi^2 (\vek^2_\bot + M_N^2
(x_i-\nu_i)^2)$ and inserted in (\ref{17}) to produce the pdf of the
multihybrid state. Integrating over $d\vek_{\bot i}~ dx_i$ in (\ref{17}) one
finally obtains with the notations $\kappa^2 = 2 \alpha, M_0 \to M_N$ the form
of pdf suggested and studied in \cite{12} \be  f_i^{(N)} (M_N|x-\nu|) =
\xi^{(N)}_i \frac{M_N}{\kappa} \exp \left(-\frac{M^2_N}{\kappa^2}
(x-\nu_i)^2\right)\label{a1}\ee where $i=g,q ; \xi_i^{(N)}$ is defined by
normalization condition

\be \int^1_0 f_i^{(N)} (x) dx = 1\label{a2}\ee and we have introduced effective
masses of quarks and gluons $ M_N= 3\omega +N m, $ $3\omega=M_p$, so that \be
\nu_g = \frac{m}{M_N}, ~~ \nu_q = \frac{\omega}{M_N}. \label{a3}\ee

Using (\ref{a1}) one can proceed as in  \cite{12}  to calculate the pdf's of
gluons $g(x)$ and valence $u$ quarks $u(x)$, and we shall neglect as in
\cite{12} the contribution of the  nucleon itself, so that in multihybrids
$N\geq 1$, and   concentrate on the region of small $x$, $x\la 0.1$. One has
for a sequence of $N$- multihybrids with  with  probabilities $|C_N|^2$

\be  u_v (x) = 2 \sum_{N=1}|C_N|^2 \xi_q^{(N)} \frac{(M_p+m_N)}{\kappa} \exp
\left( -\frac{M^2_N}{\kappa^2} (x-\nu_q)^2\right)\label{a4}\ee

\be  g_v (x) =  \sum_{N=1 }^\infty |C_N|^2 \xi_g^{(N)} N\frac{M_p+N_m}{\kappa}
\exp \left( -\frac{M^2_N}{\kappa^2} (x-\nu_g)^2\right).\label{a5}\ee

In what follows we shall slightly generalize the discussion in \cite{12} and
assume that $|C_N|^2$ decrease as $N^{-\gamma}, 1<\gamma<2$,

\be \sum|C_N|^2 =1, ~~|C_N|^2 = N^{-\gamma} |\bar c|^2, ~~ |\bar c|^2 =
\zeta(\gamma),\label{a6}\ee where $\zeta (\gamma)$ is the Riemann
dzeta-function,  $\zeta \left(\frac32\right) = (2,62)^{-1}$.

As a result one obtains for    $g(x)$ and $u(x)$

\be g(x) = |\bar c|^2 (a_1  x^{\gamma-2} + a_2 x^{\gamma-3}),\label{a7}\ee

\be u_n(x) = |\bar c|^2 (b_1 x^{\gamma-1} + b_2 x^{\gamma-2} +
b_0),\label{a8}\ee where $a_i, b_i$ depend on  parameters $\kappa, m (\omega=
\frac{M_p}{3} =0.31$(GeV). We can  fix  these parameters, as in \cite{12},
using the explicit form of the lowest hybrid with $N=1$ from \cite{16,17},
$\kappa=\frac{m}{2}= 0.313$ GeV.

In this way we have  the only fitting parameter $\gamma$, which we can choose
to  correspond roughly to the gluon pdf at  $Q^2=(3\div 10)$ GeV$^2$. In this
way  we can consider $g(x)$ and $u(x)$ as starting   pdf in  the DGLAP
evolution equations instead of standard initial state functions with roughly
15-20 free parameters. Following \cite{12} we take $\gamma =3/2$ and obtain \be
a_1=1.5, ~~a_2 =1, ~~ b_1 = 2.82, b_2=2.70 .\label{a9}\ee

As a result one obtains $xg(x)$ at the points $ x=10^{-2}; 10^{-1}$ to be equal
to 3.87; 1.38, which is close to the PDG data at $Q^2=10$ GeV$^2$, respectively
5; 1.5 (cf. Fig 19.4 in the last ref. of \cite{5,5a,5b}).

The same type of similarity between (\ref{a8})  and  and PDG data exists for
$xu_v(x)$, which implies, that the sequence of  multihybrid  states can be used
as a part of evolution process. It is also interesting that $g(x)$ contains an
additional power of $x^{-1}$ as compared to $u_v(x)$, which also nicely
corresponds to the data. One can now check whether multihybrids can produce
reasonable amount of sea  quark  pdf. This check was done in \cite{12}, where
it was  shown that   the $q\bar q$ DGLAP evolution of (\ref{a7}) in the
interval (1-10) GeV$^2$ produces $\bar u(x, 10$ GeV$^2)\approx 0.04 g (x, 10 $
GeV$^2)$, in good agreement with  PDG data \cite{5}.

As a final  check we compute $g(x, Q^2 =10^4$ GeV$^2)$, as a result of the
DGLAP evolution of our form (\ref{a7}), (\ref{a9}) for $Q^2=10$ GeV$^2$ and
obtain \be \frac{dg}{d\ln Q^2} =\frac{3\alpha_s}{\pi} \varphi (x),  \varphi(x)
\simeq |\bar c |^2 (0.917 x^{-3/2} + x^{-1}-3.4 x^{-1/2})\label{a10}\ee which
yields    roughly $xg(x=0.01, 10^4$ GeV$^2 ) \approx  3.5$, whereas the PDG
data gives a 2.5 larger value. This might point out to an additional mechanism
(e.g. of the BFKL type) at very large $Q^2$ and small $x$.

We have considered above only the region of small $x$, $x<0.1$. Larger $x$ need
some modifications. Namely, introducing the quark-counting rule factor
$(1-x)^{\rho(N)}$ into (\ref{a4}), (\ref{a5}), where $\rho(N)$  is  growing
with $N$, one obtains modified normalization factors  $\xi_i^{(N)}$ and an
effective upper limit in the  summation over $N$, $N_{\max}\sim \frac{Q}{m}
\sqrt{1-x}$, since $M^2_N (eff) \sim s$. These  replacements do not essentially
change our results for $N_{\max}\gg 1$, but  are  essential for very small $x$,
where the contribution of the multihybrid contribution is strongly damped by
$\rho(N)$ as compared to the nongluonic excited nucleon states.

 Concluding this section, we have shown
that the multihybrid mechanism at the initial or intermediate stage of parton
evolution can be considered on the same ground as other realistic contributions
and may have its own part in the final reaction products.

Note, that our small $x$ asymptotics, $u_g(x) \sim x^{-\gamma}$, is different
from the reggeon-type asymptotics, and is a result of the multihybrid state
formation in the general multigluon ladder-type diagrams. It is possible, that
the latter provide both types of asymptotics, the more common ``hard'' momentum
distribution when each loop in the ladder with a large rapidity  ratio yields
$\left(\alpha \ln \frac{1}{x}\right)$, and a soft (or coherent) momentum
distribution, as in (\ref{30}), which possibly occurs for very large $s$. We
study below in the next sections and appendix these two regimes in the example
of the square box diagram with confinement, and discover two possible regimes,
depending on whether the  external momenta are much larger, or comparable with
the string tension.

Significantly as one of possible signals one can consider the ridge-type
configurations \cite{27,27a,27b}, where experimental signals are along a
straight line, which mimicks the average form of a large multihybrid with $N\gg
1$.

\section{Nonperturbative vs perturbative regimes}

 The purpose of the present section (and the Appendix 2) is to demonstrate,
 that IR and  collinear singularities are absent in QCD (in  contrast to QED),
 which however does not imply the absence of the   logarithmic and double
 logarithmic terms in the perturbative series.

As a standard approach to the evolution and cross sections in high-energy QCD
one is using the QED processes of $e\to e'\gamma, ~~ \gamma \to  e^+ e^-$ with
the IR and collinear singularities and perform the analogous perturbative
procedure in QCD, leading to the  singularities e.g. in  fragmentation
functions and average number of partons etc. However, in QCD both IR and
collinear singularities are absent, as we demonstrate below, and the
multiplicative gluon emission leads to the formation of heavy multihybrid
states, in contrast to the QED, where the multiphoton emission is a real
background process,  which is treated introducing an experimental bound -- the
minimal energy $E_l$ of the registered photons \cite{21,21a,21b}.

As a consequence  in  QED the observed cross section of  the high $q^2$ process
contains Sudakov double-logarithmic corrections of the type $\frac{\alpha}{\pi}
\ln \left( \frac{q^2}{m^2} \right) \ln \left( \frac{q^2}{E^2_l}\right).$

In QCD the minimal energy $E_l$ is absent, since gluons are never emitted as
free particles and its emission inside hybrid states costs an increase in the
total mass  of ground states $\Delta E \approx O(1$ GeV). Moreover, the quark
mass $m_q$ cannot enter in the final equations due to confinement for $m_q<
\sqrt{\sigma} =0.42$ GeV, hence the multigluon emission might have another
structure. In the DGLAP evolution equations this difficulty is avoided
considering evolution of quark, antiquark and gluon pdf's $f_f (x,Q)$ as
functions of momentum scale $Q $,  with the kernels ensuring logarithmic
dependence on $Q^2$ (a weak violation of the Bjorken scaling), and on $x$ at
small $x$.

As we argue below in this section, and in the Appendix 2,  in QCD the IR and
collinear singularities  are absent, but the perturbative series terms can
indeed have the known logarithmic form, and  the role of the  cut-off parameter
is played by the string tension, so that at high momenta, $p^2\gg \sigma$, the
perturbative kinematics prevails.

We start with the standard perturbative picture. To put the problem in the most
simple form, consider the process $e^+e^-\to q\bar q g$, with the cross section
for free $q, \bar q, g$ equal to (see e.g. \cite{4}, chapter 17) \be
\frac{d\sigma}{dx_1 dx_2} (e^+e^- \to q\bar q g) = \sigma_0 \left( 3 \sum_f
Q_f^2\right) \frac{2\alpha_s}{3\pi} \frac{x_1^2+x_2^2}{(1-x_1)
(1-x_2)}\label{31}\ee where $x_1, x_2, x_3$ are ratios of $q, \bar q$ and $g$
energies to the $e^+e^-$ energy $\sqrt{s}$ and one can write \be 1-x_1 = x_2
\frac{E_g}{\sqrt{s}} (1-\cos  \theta_{2g}), ~~ 1-x_2 = x_1 \frac{E_g}{\sqrt{s}}
(1-\cos \theta_{1g}),\label{32}\ee
 and these denominators arising from the virtual propagators of quarks:\\
 $(p'_1)^{2} - m^2_1=  (p_1 + p_g)^2 = m_1^2 +\sqrt{s} E_g x_1 (1-\cos_{1g} ) $.
   One can see the collinear and IR singularities of the unconfined
 QCD, which are essential element in the  balance of the integrals defining the terms of the perturbative series.

\begin{center}

\begin{figure}
\begin{center}
  \includegraphics[width=10cm, ]{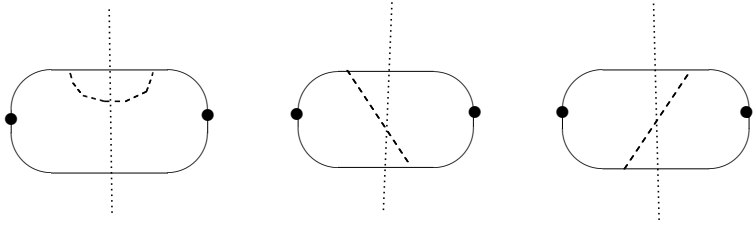}

\end{center}
  \caption{The Feynman diagrams for the $q\bar q$ Green's function,
   which contribute to the $q\bar q g$ cross production cross section}

\end{figure}
 \end{center}

 In general, the bremsstrahlung probability can be  written as

\be  dP_{brems} = \frac{\alpha C_N}{\pi^2} \frac{ d^2 k_\bot}{k_\bot^2}
\frac{dx}{x},\label{33s}\ee where $x\sim \frac{m^2}{s}$, or else $x\sim
\frac{Q^2}{s}$ for the deep inelastic evolution. In the series of the gluon
cascade amplitude (the Regge-Gribov amplitude) one has for the term with $n$
internal lines

\be I_n \sim (c\alpha_s)^n \int^1_x \frac{dx_n}{x_n} \int^1_{x_n}
 \frac{dx_{n-1}}{x_{n-1}}... \int^1_{x_2} \frac{dx_1}{x_1} \approx \frac{1}{n!}
 ( c \alpha_s \ln \frac{1}{x})^n\label{34s}\ee
 which yields for the total sum the standard answer
\be \sum_n I_n \sim \exp {(c\alpha_s \ln \frac{1}{x})} \sim\left(
\frac{s}{m^2}\right)^{c\alpha_s}.\label{35s}\ee

One can see, that the crucial property of the Regge-type behavior (\ref{35s})
is the bremsstrahlung-type energy distribution  $\frac{dx}{x}$, at each step of
the gluon cascade, as it is also in (\ref{31}). Therefore it is  interesting to
calculate the same cross section (\ref{31}) with confinement taken into
account.

 To  make a comparison with  the $np$ (confined) QCD,   one must
 calculate the diagrams, shown in Fig.2, and one has in mind, that all the area
 inside the outer contour  of the diagrams, is covered with the confining film
 (shown by the vertical lines in Fig. 3).

\begin{center}

\begin{figure}
\begin{center}
  \includegraphics[width=10cm, ]{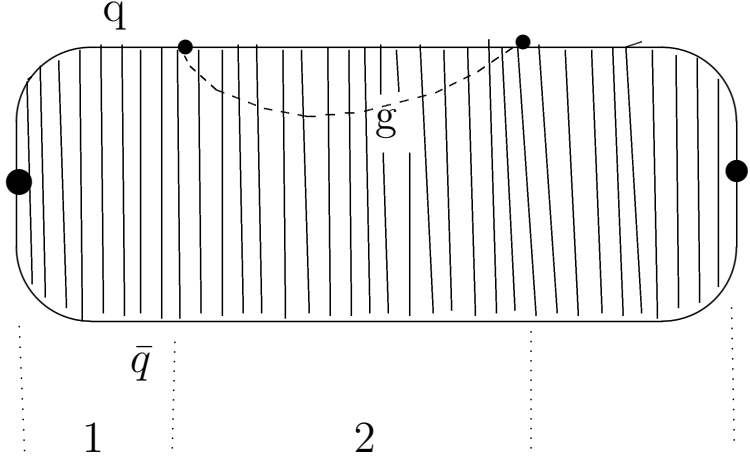}

\end{center}
  \caption{The  path integral representation of the  $q\bar q$ Green's
  function. The regions denoted by 1,2,3 refer to $q\bar q, q\bar q g$ and
   $q\bar q $  intermediate states respectively}

\end{figure}
 \end{center}

  We calculate the  $q\bar q$ Green's function in the path integral formalism
  incorporating confinement and gluon exchange in appendix 1 and write down the
  result in the  form, appropriate  for the Fock component language, namely
  \be G_{q\bar q} (E) = \frac{\lan Y\ran}{(2\omega)^2} \varphi (0)
  \frac{1}{E-E_1} \frac{V_{12}}{\sqrt{2\omega_g}} \frac{1}{E-E_2}\frac{V_{23}}{\sqrt{2\omega_g}}
   \frac{1}{E-E_3} \varphi (0)\label{33}\ee
   where the gluon creation matrix elements $V_{ik}$ are expressed via overlap
   matrix elements of meson and hybrid wave functions \cite{14,15,22,22a}, and
   $\omega, \omega_g$ are average energies (from the Hamiltonian minimization)
   of quarks and gluon.

   From \cite{15} one has
   \be \frac{V_{12}}{\sqrt{2\omega_g}}=\frac{V_{23}}{\sqrt{2\omega_g}}\cong g
   \cdot 0.08 {\rm GeV}\label{34}\ee

 To obtain the cross section $ e^+e^- \to hybrid (q\bar qg)$, one takes the discontinuity  of
 (\ref{33}) in  energy due to the factor $\frac{1}{E-E_2}$ and obtains
 \be d \sigma (e^+e^- \to hybrid) \sim \left[ \varphi(0) \frac{V}{\sqrt{2\omega
 g}} \frac{1}{E_2 -E_1}\right]^2 \delta (E-E_2)\label{35}\ee
 where $E_2-E_1\approx 0.8 \div 1 $ GeV for $E_1 \approx m_\rho, ~~ E_2 =
 M_{hybrid}$ and one does not have any IR or collinear singularities in
 agreement with our discussion and the beginning of this section. It is also
 essential, that the structure of the  final states in (\ref{35})  in the  np regime is quite
 different from the  perturbative QCD, namely for low energy resolution one
 should sum up over all collective states of the hybrid, and in the case of
 high energy and multihybrid
 state the overall entropy of states can  be very large.

To   resolve the apparent contradiction one can consider the Feynman diagrams,
shown in Figs. 4, 5,  where we take  into account confinement  as  the area law
of the
  Wilson loop, e.g. formed by the 4 sides of the rectangular in Fig. 5.

\be W(C) = \exp (-\sigma S_{min} (1,2,3,4))\label{36s}\ee

As shown in the appendix 2, the approximate answer for this diagram is \be  G_4
(p_i) =   \prod^4_{i=1} \int d^4 q_i  \int^\infty_0 ds_i e^{-s_i (m^2_i
+q^2_i)} J_\sigma (s_i, b^{(i)})\label{37s}\ee and \be
 b^{(i)} = p_i - q_i + q_k\label{38s}\ee
where $q_1, q_k$ are momenta on the adjacent lines in the vertex $i$. The
property of $J_\sigma$ is such, that at large $p_i, q_k, (b^{(i)})^2 \gg
\sigma$, it becomes a product of the vertex $\delta$- functions

\be J_\sigma ((b^{(i)})^2\gg\sigma) \sim  \prod^4_{i=1} \delta^{(4)} (b^{(i)})
\label{39s}\ee

As a result  $G_4(p_i)$ coincides in this limit with the standard perturbative
answer, where the bremsstrahlung property (\ref{33s}) appears, but now for
large (as compared with $\sqrt{\sigma}$) momenta.

At the same time for low momenta, $(b^{(i)})^2\la\sigma$, one has four $d^4q$
integrations, which  exclude any possibility of IR  or collinear singularities.

Thus one may imagine for a sequence of ladder-type diagrams two types of
asymptotics, depending on the internal momenta distributions: a) (\ref{34}),
(\ref{35}) for  perturbatively ordered momenta; b)multihybrid-type asymptotics
considered in the  previous section for np ordered momenta, $ |\Delta P_i| \sim
\sqrt{\sigma}$.

 We are now coming to a possible general picture of np evolution in DIS or
 high energy hadron reactions. We assume,
 that in the  intermediate or initial state there appears a multihybrid state (or a family of
 multihybrid states with close-by masses).
 This multihybrid state originally (by evolution) is perturbatively (bremsstrahlung-type) ordered, which corresponds to the  reggeon-like
 asymptotics and  the BFKL-type desctiption, but has a possibility to develop
 into a quark stable multihybrid state, described in the previous section. It   subsequently decays into more
 multihybrids and finally  into $q\bar q$ and $3q$ hadrons, due to perturbative and
 np string breaking. In this way the gluons  contribute  the  most amount of the collision energy at this  primordial
 stage, which in the evolution process is transferred to the final hadrons.

\section{Summary and discussion}

We have tried above to  formulate main features of the  np approach to the high
energy production of hadrons in DIS or hadron-hadron collisions, which might be
an alternative or a complement  to the standard  QCD picture \cite{5,5a,5b}.
The basis of our approach is the fact, that the Lorentz contraction rule of
wave functions automatically leads to the parton-like form of  its momentum
depenence \cite{11}, and one  the c.m. wave function to rewrite it  in terms of
partonic variables, $\vep_\bot$  and $x$, namely $\vep^2 =\vep^2_\bot +M^2_0
(x-\nu)^2$, where $M_0$ is the c.m. energy of the object.

This allows to write down the pdf's of the object, e.g. of the fast proton in
terms of the c.m. wave function, as it was done in \cite{12} both in  the
polarized and  unpolarized cases.  In this way one obtains the np components of
the partonic set of pdf's, and the point is how to represent wave function of
the arbitrary moving complex object and subsequently how to extract from it the
total set of pdf's: valence and sea for any quark flavor.

This is done at the beginning of the present paper with the help of Fock
components of the Fock column wave function, where each line corresponds to a
definite ensemble of $q ,\bar q, g$  with the same total quantum numbers. We
have expressed pdf's as a sum over Fock  components, satisfying usual
normalization conditions. We have also shown before in \cite{12}, that the
normalization of the Fock components is boost invariant, and hence it is
possible and often convenient to calculate pdf's from the Fock column in the
rest frame.

Using that, we have considered in section 3 the basic building block of the
high excited Fock column -- the  multihybrid state, which may be one of  the
basic primordial state of the high energy evolution of the Fock column in the
$1/N_c$ expansion, where the $g \to q\bar q$  decays are $O(1/N_c)$, as
compared to gluon creation and absorption.

We have found   the wave function of the multihybrid and discovered that it is
highly correlated with an average energy per quark and gluon around 0.6 GeV and
average intergluon momentum $|\Delta \vep_\bot|$ around 0.36  GeV, and $|\Delta
x_i|\sim \frac{1}{3.5N}$, where $N$ is the total number of particles.

Such  a complicated object for large $N$ has a very large enthropy and can be
stable enough with respect to emission and reabsorption of quarks and gluons,
and hence can  be a prototype of the structures of the ridge type, observed in
$pp$ and $AA$ collisions \cite{24,24a}, \cite{25a}.

We have  also shown, that the appearance  of a multihybrid at some stage of the
evolution does not violate the observed pdf's of gluons, valence quarks and sea
quarks, however at very large $E^2 \sim Q^2/x$ one  possibly needs combined
effects of both reggeons and multihybrids.

 Moreover, an ensemble of
multihybrids yielding a balance of perturbative and np features different from
the standard approach \cite{25b, 25c}, can be a reasonable alternative
saturation state of the system evolution ( see \cite{25} for a discussion),
since for a given excitation energy $E$ of the system  the set of equations for
Fock components, leads to the effective cut-off of the number $N$ of
constituents, $ N\approx \frac{E}{\varepsilon_g} \approx \frac{E}{0.8~ {\rm
GeV}}$ and this set is subsequently hadronized. Finally, one should stress some
similarity between the multihybrid mechanism and the Lund string model
\cite{26}. We note also, that  the decay of a multihybrid can proceed or via
gluon decay $g\to q\bar q$, or else via np decay of the string piece into
$q\bar q$, which initiate different genealogical chains. Note also, that the
strong np decay into $q\bar q$ pair can be accompanied by pion emission, -- the
phenomenon observed e.g. in the heavy quarkonia \cite{27,27a,27b}, which can
produce pions at the earliest stages of evolution.

The author is  grateful  for useful discussions to   K.G.Boreskov, B.L.Ioffe,
O.V.Kancheli and members of  the ITEP theory seminar and to  I.M.Dremin and
members of the FIAN theory seminar. The author gratefully acknowledges the help
of I.V.Musatov in  working out the material of appendix 2. The financial
support of the RFBR grant 1402-00395 is gratefully acknowledged.

\setcounter{equation}{0} \def\theequation{A1.\arabic{equation}}

\vspace{2cm}
 \setcounter{equation}{0}
\renewcommand{\theequation}{A.\arabic{equation}}

\hfill {\it  Appendix  1}

\centerline{\it\large Calculation of the $q\bar q g$ Green's function in the
path integral formalism }

 \vspace{1cm}

\setcounter{equation}{0} \def\theequation{A1.\arabic{equation}}

We start with the path integral form of the free Green's function

\be  g(x,y) = \left( \frac{1}{m^2- D^2 } \right)_{xy} = \sqrt{\frac{T}{8\pi}}
\int^\infty_0  \frac{d\omega}{\omega^{3/2}} (D^2 z)_{\vex\vey} e^{-K(\omega)}=
\sqrt{\frac{T}{8\pi}}\int \frac{d\omega}{\omega^{3/2}} \lan \vex |e^{-H(\omega)
T}|\vey\ran\label{A1.1}\ee where $T=x_4-y_4$ and \be K(\omega) = \int^T_0 dt_E
\left( \frac{\omega}{2}+ \frac{m^2}{2\omega} +
\frac{\omega}{2}\left(\frac{d\vez}{dt_E}\right)^2\right),~~ H(\omega) =
\frac{\vep^2+m^2}{2\omega} + \frac{\omega}{2}.\label{A1.2}\ee

In a similar way for the product of two spinor $q, \bar q$ Green's function one
has

$$  G_{q\bar q}(x,y) = \left( \frac{(m_1-\hat D_1)(m_2-\hat D_2)}{(m^2_1- \hat
D^2_1)(m^2_2-\hat D^2_2) } \right)_{xy} = {\frac{T}{8\pi}} \int^\infty_0
\frac{d\omega_1}{\omega^{3/2}_1} \int^\infty_0
\frac{d\omega_2}{\omega^{3/2}_2}\times$$\be \times (D^3 z_1)_{\vex\vey} (D^3
 z_2)_{\vex\vey} e^{-K_1(\omega_1)- K_2(\omega_2)}\lan YW \ran
\label{A1.3}\ee where \be  4Y = tr~ \Gamma (m_1-\hat D_1) \Gamma(m_2-\hat
D_2);~~ W= tr\exp ig \int_C A_\mu dz_\mu\label{A1.4}\ee
 and we have omitted for simplicity the spin-dependent terms in $W$. To
 implement the gluon lines in the total amplitude, we can write the
 path-integral form
 \be \lan A_\mu (u) A_\nu (v) \ran = \delta_{\mu\nu} \int d s e^{-K}
 (D^4z)_{uv}\Phi(u,v)\label{A1.5}\ee where $\Phi(u,v) = P\exp (ig \int^u_v
 A_\lambda dz_\lambda)$.

 We also take into account the identities  \cite{8,8a,8b,8c,8d,8e,8f}, last reference,
 \be (D^4z)_{xy} = (D^4z)_{xu} d^4u (D^4z)_{uy}\label{A1.6}\ee

\be \int^\infty_0 ds \int^s_0 d\tau_1\int^{\tau_1}_0 d\tau_2 f(s,\tau_1,
\tau_2)= \int^\infty_0 ds \int^\infty_0 d\tau_1\int^{\infty}_0 d\tau_2
f(s+\tau_1+ \tau_2, \tau_1+ \tau_2, \tau_2).\label{A1.7}\ee

Our purpose is  to write the path integral representation of  the diagram in
Fig.1. Using (\ref{A1.3}), (\ref{A1.6}) and (\ref{A1.7}) one can write $$
G_{q\bar q}(x,y) = \int ds_1 d\tau_1 da_1 ds_2  d\tau_2 da_2 (D^4 z)_{xu} d^3u
2\omega_{12} (D^4z)_{uv} d^3v 2\omega_{23}\times$$ \be \times (D^4z)_{vy} \hat
T W_\sigma e^{-K_1-K_2} (D^4z')_{xu'} d^3u' 2 \omega'_{12}(D^4z')_{u'v'}d^3v'
2\omega'_{23} (D^4z)_{v'y}\int ds'~ e^{-K_g}(D^4z)_{uv} W_A.\label{A1.8}\ee

 One can now use the correspondence in (\ref{A1.1}) to introduce the
 Hamiltonians $H_1,H_2, H_3$ for three sectors 1,2,3 shown in Fig.1, namely
\be \int ds_1 (D^4z)_{xu} e^{-K_1} = \sqrt{\frac{t}{8\pi}} \int
\frac{d\omega_1}{\omega_1^{3/2}} \lan \vex |e^{-H_1t}|\veu\ran\label{A1.9}\ee
and similarly for $H_2$, $H_3$. As a result the total $q\bar q$ Green's
function after integrating out the c.m. coordinates is

$$ G_{\vep} (T) = \int G_{q\bar q} (x,y) e^{i\vep(\vex-\vey)} d^3 (\vex-\vey) =
$$ $$ = \frac{\lan Y\ran}{4} \int \frac{t}{2\pi} \frac{d\omega_1
d\omega_2}{(\omega_1 \omega_2)^{3/2}} \lan 0| e^{-H_1t}|\verho_u\ran
d^3\verho_u 2\omega_{12} 2\omega'_{12} \frac{t'-t}{8\pi}
\frac{d\omega'_1d\omega'_2}{(\omega'_1\omega'_2)^{3/2}}\times$$
$$\times \sqrt{\frac{t'-t}{8\pi}} \frac{d\omega'_3}{\omega_3^{\prime 3/2}} \lan
\verho_u, \verho'_u\ e^{-H_2 (t'-t)}|\verho_v, \verho'_v\ran d^3\verho_v
2\omega_{23} 2\omega'_{23}\times$$ \be \times \frac{T-t'}{8\pi} \int
\frac{d\omega^{\prime\prime}_1
d\omega^{\prime\prime}_2}{(\omega^{\prime\prime}_1\omega^{\prime\prime}_2)^{3/2}}
\lan \verho_v | e^{-H_3 (T-t')}|0\ran dtdt'\label{A1.10}\ee and finally,
integrating over Euclidean time intervals, one obtains \be G_{q\bar q} (E) =
\int G_{\vep} (T) e^{iET} dT= \frac{\lan Y\ran}{(2\omega)^2} \varphi_1 (0)
\frac{1}{E-E_1} \frac{V_{12}}{\sqrt{2\omega_g}}\frac{1}{E-E_2}
\frac{V_{23}}{\sqrt{2\omega_g}}\frac{1}{E-E_3}\varphi_2 (0)\label{A1.11}\ee
where $V_{ik}$ are defined as (see also \cite{15}) \be V_{ik} \equiv
V_{ik}^{(\mu)} = g \int \varphi_M (\ver)~^\mu \psi^+_n (0,r) d^3r,
\label{A1.12}\ee where $\varphi_\mu^{(k)} (\ver)$ and $~^\mu\psi_n (\ver_{12},
\ver_{23})$ are wave functions of the meson in the sector 1 or 3,  and
respectively of the  hybrid with $\mu$ -- the gluon spin orientation.

The form (\ref{A1.11}) is exactly what is expected in the Fock component
formalism of section 2.


\vspace{2cm}
 \setcounter{equation}{0}
\renewcommand{\theequation}{A.\arabic{equation}}

\hfill {\it  Appendix  2}

\centerline{\it\large Integral representation for the 3 and 4 point functions}

 \vspace{1cm}

\setcounter{equation}{0} \def\theequation{A2.\arabic{equation}}

In this Appendix  we shall consider the path integral technic  \cite{9,9a} for
3 and 4-point amplitudes with one closed quark contour and any number
 of perturbative and nonperturbative gluon interactions.
The basic technic was described in \cite{9,9a},  and  we start with  the
Euclidean space-time. Then  the $n$-point amplitude defined as
 \be
G(p_i^{(1)},...p_n^{(n)})=\lan J_1(p_1)... J_n(p_n)\ran,~~ J_i(x) =\bar\psi(x)
\Gamma_i\psi(x)\label{x1} \ee can be written using \cite{9} as \be
G(p_n^{(1)}...p_k^{(n)})= \lan tr \prod^n_{i=1} \Gamma_i(m_i-\hat D_i)
\int^\infty_0 ds_i(Dz^{(i)})_{x^{(i)},x^{(i-1)}} e^{-K_i} \Phi^{(i)}_ \sigma
e^{ip^{(i)}x^{(i)}} dx^{(i)}\ran.\label{x2} \ee Here we have denoted \be
\Phi_\sigma^{(i)}(x^{(i)},x^{(i-1)})= P_AP_F\exp(ig \int^{x^{(i)}}_{x^{(i-1)}}
A_\mu dz_\mu) \exp (g\int^{s_i}_0 d\tau_i \sigma_{\mu\nu} F_{\mu\nu}),
\label{x3} \ee
$$\sigma_{\mu\nu} \equiv \frac{1}{4i}
(\gamma_\mu\gamma_\nu-\gamma_\nu\gamma_\mu),$$ and \be (Dz^{(i)})_{xy}
=\lim_{N\to\infty}\prod^N_{k=1}
\frac{d^4\xi^{(i)}(k)}{(4\pi\varepsilon)^2}\frac{d^4q^{(i)}} {(2\pi)^4}
e^{iq^{(i)}(\sum_k\xi^{(i)}(k)-(x-y))},~~ N\varepsilon =s. \label{x4} \ee
Eq.(\ref{x2}) is the exact expression in QCD,
 when no internal quark loops
 are present, so it should be exact in the large $N_c$ limit.

All phase factors (\ref{x3})   combine in (\ref{x2}) into  a Wilson loop
 factor with insertion of operators $\sigma F$,
which we denote as $\lan W_\sigma\ran$, \be \lan W_\sigma\ran = \lan
\prod^n_{i=1} \Phi_\sigma^{(i)} \ran_A. \label{x5} \ee
 In what follows we disregard the spin factors $\sigma F$, in the
 first approximation
since as it was shown  in \cite{8,8a,8b,8c,8d,8e,8f}, they give nondominant
contribution to the amplitude at large momenta. Moreover the $(m_i-\hat D_i)$
factors when acting on path integrals in (\ref{x2}) are shown to be written as
$(m_i-i\hat q_i)$,
   where $q_i$ are Minkowskian
 momenta on the given line (see  last reference in \cite{8,8a,8b,8c,8d,8e,8f} for
a derivation).

\begin{center}

\begin{figure}
\begin{center}
  \includegraphics[width=10cm, ]{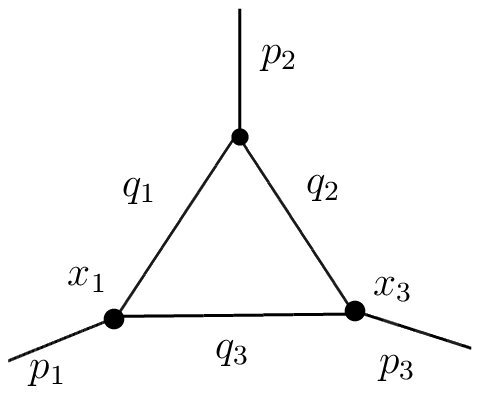}

\end{center}
  \caption{}

\end{figure}
 \end{center}

We start here to calculate $\lan W_\sigma\ran$ for the triangle diagram of
Fig.5, making the approximation of large area, $S_{min} \gg T^2_g$, so that \be
\lan \lan W_\sigma\ran_A\approx \exp (-\sigma S_{min})\label{x11} \ee
 Here
$S_{min}= S_{min}(x^{(1)}, x^{(2)}, x^{(3)}$) and we assume for simplicity the
straight-line (eikonal) geometry for the quark lines, so that the Nambu-Goto
expression  for $S_{min}$ can written as \be S_{min}= \int^1_0 dt \int^1_0
d\beta \sqrt{\dot{w}^2 (w ')^2- (\dot w w')^2},~~ \dot w \equiv\frac{\partial
w}{\partial t}, w'\equiv \frac{\partial w}{\partial \beta} \label{x12} \ee and
quark trajectories $\bar z_\mu$  (from $x^{(1)}$ to $x^{(3)}$)
 and $z_\mu$ (from $x^{(1)}$ to $x^{(2)}$)
are given as $$ z_\mu(t) =x^{(1)}_\mu+ (x^{(2)}-x^{(1)})_\mu t $$ \be \bar
z_\mu(t) =x^{(1)}_\mu+ (x^{(3)}-x^{(1)})_\mu t\label{x13} \ee with \be
w_\mu(\beta, t) =z_\mu(t) \beta+\bar z_\mu(t) (1-\beta).\label{x14} \ee

Since the Nambu-Goto form (\ref{x12}) is difficult to handle,
 one introduces as usual einbein variables
$\nu(\beta,t)$ and $\eta (\beta,t)$ to write \be e^{-\sigma S_{min}} = \int
D\nu D\eta e^{-{2}\int\nu [\dot w^2 +(\frac{\sigma}{\nu})^2w^{\prime
2}-2\eta(\dot w w')+ \eta^2(w')^2]d\beta dt}.\label{x15} \ee

The form in the exponent in (\ref{x15}) is quadratic in coordinates $x^{(i)}$
and can be easily computed to be \be e^{-\sigma S_{min}} = \int D\bar\nu D\eta
\exp\{-\sigma \frac{\bar \nu}{4} f\}, \label{x16} \ee where \be f\equiv
y^2_1+\frac{1}{\bar \nu^2}y^2_2-2\eta y_1y_2+\eta^2 y^2_2,\label{x17}\ee
 and we have used notation $\bar \nu = \nu/\sigma $ and
\be y_1\equiv x_{12} =x^{(1)} -x^{(2)};
 y_2\equiv x_{13} =x^{(1)} -x^{(3)}.\label{x18}
\ee The full integral over $x^{(i)}$ which is contained in (\ref{x2}) can be
written as $$ I_3 \equiv \int dx^{(2)} e^{i(p^{(1)}+p^{(2)}+ p^{(3)})x^{(2)}}
dy_1 dy_2 e^{i(p^{(1)}+p^{(3)} )y_1-ip^{(3)}y_2}\times$$ \be \times
e^{iq_1y_1+iq_2(y_2-y_1)-iq_3y_2}\exp\{-\sigma \frac{\bar \nu}{4} f\} d\nu
d\eta.\label{x19} \ee

\begin{center}

\begin{figure}
\begin{center}
  \includegraphics[width=10cm, ]{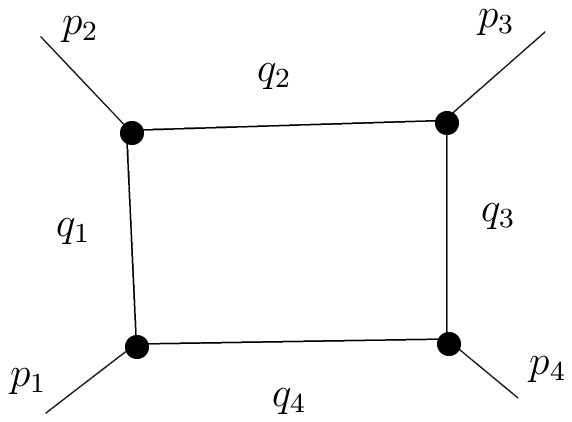}

\end{center}
  \caption{}

\end{figure}
 \end{center}

The  coordinate part of the integration in (\ref{x19}) can be written as \be
J_3(a,b)\equiv \int d^4 y_1 d^4 y_2 e^{-(a_{ik}y_iy_k+b_iy_i)}=
\left(\frac{\pi}{\det a}\right)^2 e^{-\frac{1}{4\sigma} b_i (a_{ik})^{-1}
b_k}\label{x20} \ee where notations are used $$ a_{11}=\frac{\bar \nu}{4};
~~b_1=-(p^{(1)}+p^{(3)})-q_1+q_2 $$ $$
a_{22}=\frac14(\frac{1}{\bar\nu}+\eta^2\bar \nu); ~~ b_2=p^{(3)}-q_2+q_3 $$ \be
a_{12}=a_{21} =-\frac{\eta \bar \nu}{4}. \label{x21} \ee Inserting (\ref{x20})
into (\ref{x19}) one has \be I_3=(2\pi)^4 \delta (\sum p^{(i)}) \int D\bar
\nu\int D\eta \left( \frac{\pi}{\det a}\right)^2 \exp \left(-\frac{1}{4\sigma}
b_ia_{ik}^{-1} b_k\right).\label{x22}\ee Since $\nu,\eta$ are nondynamical
variables, the integrals over them can be
 taken using the stationary point analysis, which amounts to
solving equations \be \frac{\delta
\Psi_3}{\delta\eta}\left|_{\eta=\eta_0}=0,\right.~~ \frac{\delta
\Psi_3}{\delta\bar \nu}\left|_{\bar \nu=\nu_0}=0,\right.\label{x23} \ee with
\be \Psi_3 = -2\ln \det a +\frac{1}{4{\sigma}}  b_{i} a_{ik}^{-1}
b_k.\label{x24} \ee Inserting $\bar \nu=\nu_0(p^{(i)}, q_i),~\eta=\eta_0
(p^{(i)}, q_i)$ back into (\ref{x22}) one has $I^{(0)}_3(p^{(i)}, q_i)$ and the
3-point amplitude acquires the form \be G(p^{(1)},p^{(2)}, p^{(3)}= tr
\prod^3_{i=1}\int d^4q_i \Gamma_i (m_i-i\hat q_i)\int^\infty_0 ds_1
e^{-s_i(m^2_1+q^2_i)} I_3^{(0)}(p^{(i)}, q_i),\label{x25}\ee \be I_3^{(0)} =
(2\pi)^4 \delta (\sum p^{(i)}) (16 \pi)^2 \exp (-\frac{2}{\sigma}
\sqrt{b^2_1b_2^2- (b_1b_2)^2}).\label{x25a}\ee
 The exponential
factor in $I_3^{(0)}$ displays the role of confinement on the 3-point Green's
function. Note first of all, that $b_1, b_2$ defined in (\ref{x21}) and $b_3=
p^{(1)} +q_1-q_3$ satisfy the relation $b_1+b_2+b_3=0$ and therefore one can
check that the factor in the exponent in (\ref{x25}) is symmetric with respect
to the replacements $b_1, b_2\to b_1,b_3\to b_2,b_3;$ \be \tilde S (b_1,
b_2)\equiv \sqrt{b_1^2b_2^2-(b_1b_2)^2}= \tilde S(b_1, b_3) = \tilde S (b_2,
b_3). \label{x27a}\ee Secondly, the magnitude of $b_i$ is the measure of "the
off-shellness" in the corresponding vertex.

In the absence of confinement $J_3(a,b)$, Eq.(\ref{x20}) goes over into the
momentum  conservation at each vertex, \be J_3(a,b)|_{\sigma\to 0} = (2\pi)^8
\delta^{(4)} (b_1) \delta^{(4)} (b_2).\label{x28a}\ee

For nonzero $\sigma$ the role of a smeared $\delta$-functions is played by the
exponential factor in (\ref{x25a}). It is clear that for large $p^{(i)},
(p^{(i)})^2\gg\sigma$, this factor strongly cuts  off all configurations of
momenta unless $b^2_i\la \sigma, i=1,2$. This means that in the presence of
confinement the momentum conservation at each vertex is fulfilled only with the
absolute accuracy of the order of $\sigma$, i.e. $q_i-q_j=
p^{(i)}+0(\sqrt{\sigma})$.

As a result the Feynman rules for the triangle graph with the area law change
in such a way that instead of one momentum integration inside the loop one has
to do all three integrations over virtual momenta $q_i, i=1,2,3$ with the
cut-off factor (\ref{x25a}). This fact strongly changes the IR properties,
since IR divergencies are now impossible by a simple power counting.

On the other hand, for large momenta, $ (p^{(i)})^2\gg \sigma, ~ q^2_i \gg
\sigma$ the exponential factor in (\ref{x25a}) strongly cuts off the  diagram
magnitude  unless only one momentum direction is available, while the
perpendicular directions are controlled by $\sigma$.
 Consider now the 4-point function.
According to the Fig.5, we assign the quark and antiquark trajectories as
follows: $$ z_\mu(t)=x_{2\mu}+(x_3-x_2)_{\mu}t $$ \be \bar
z_\mu(t)=x_{1\mu}+(x_4-x_1)_{\mu}t;~~ 0\leq t\leq 1 \label{x28} \ee and for the
standard Nambu-Goto expression (\ref{x12}) we define the straight-line metric
paths $w_\mu$: \be w_\mu(\beta, t)= z_\mu(t)\beta +\bar z_\mu(t)
(1-\beta)\label{x29} \ee we also denote three independent distances:

\be y_1=x_{12}=x_1-x_2;~~y_2=x_{23}\equiv x_2-x_3;~~y_3=x_3-x_4,~~
y_4=x_{41}=x_4-x_1;\label{x30} \ee with relation for $y_{i}$: \be \sum^4_{i=1}
y_i=0 . \label{x31} \ee Using (\ref{x15}) we define the same integral as in
(\ref{x19}) and the integral equivalent to (\ref{x20}) appears to be: \be
J_4(a,b)=\int d^4y_1d^4y_2d^4y_4\exp\{-[a_{ik}y_iy_k+b_iy_i]\}. \label{x32} \ee
It is convenient to extract string tension from $a_{ik}$, and define: \be
\nu=\sigma v, ~~ a_{ik} =\sigma \bar a_{ik}\label{x33} \ee where
$$\bar a_{11}=\frac12(\frac{1}{v}+\eta^2v); \bar
a_{22}=\frac{v}{6}+\frac16(\frac{1}{v}+\eta^2v)-\frac{\eta v}{4};
$$ $$ \bar
a_{44}=\frac{v}{6}+\frac16(\frac{1}{v}+\eta^2v)-\frac{\eta v}{4}; \bar
a_{12}=\frac14(\frac{1}{v}+\eta^2v-\eta v); $$ $$ \bar
a_{14}=\frac14(\frac{1}{v}+\eta^2v-\eta v); \bar
a_{24}=\frac{v}{12}+\frac16(\frac{1}{v}+\eta^2v); $$ $$ \bar a_{ik}=\bar
a_{ki};~~ b_1=-q_1-q_3-p_1-p_4; $$ \be b_2=q_2-q_3+p_3; ~~b_4=q_3-q_4+p_4.
\label{x34} \ee The integral $J_4$ in(\ref{x32}) is Gaussian and the result is
\be J_4(a,b)= \frac{\pi^2}{(\det a)^2} e^{\frac14 b_i(a_{ik})^{-1}b_k}.
\label{x35} \ee

In this case as also for the 3-point function, both $\eta$ and $\nu$ are to be
found from equations  equivalent to (\ref{x23}), (\ref{x24}).

One can simplify the foregoing expressions using the approximation, when the
area of the (generally speaking) nonplanar quadrangle is replaced by the sum of
areas of two triangles formed by 4 sides and one diagonal inside the
quadrangle. This approximation is exact for the planar quadrangle and gives a
larger area for a nonplanar case. It can be used to obtain a qualitative
estimate and the upper bound for the area law correction to the 4-point
amplitude.

With the notations from (\ref{x30}) one can thus write for the area of the
quadrangle, $$ S_{min}^{(4)}(1,2,3,4) \approx S_{min}^{(3)}(1,2,3)
+S_{min}^{(3)} (1,3,4)= $$ \be \frac12 \sqrt{y^2_1y^2_2-(y_1y_2)^2} +\frac12
\sqrt{y^2_3y^2_4-(y_3y_4)^2}.\label{x47}\ee The equivalent of Eq.(\ref{x19})
reads \be I_4\equiv (2\pi)^4\delta(\Sigma p^{(i)})\prod^4_{i=1}
d^4y_i\frac{d^4\mathcal{P}}{(2\pi)^4} e^{ib_iy_i-\sigma
S^{(4)}_{min}}\label{x48}\ee with $$ b_1=q_1-p_2-p_3+\mathcal{P},~~
b_2=q_2-p_3+\mathcal{P},$$ \be b_3=q_3+\mathcal{P},
~~b_4=q_4+p_4+\mathcal{P}.\label{x49} \ee Using (\ref{x47}) one obtains \be
I_4=(2\pi)^4 \delta (\Sigma p^{(i)}) \int \frac{d^4\mathcal{P}}{(2\pi)^4} \left
(\frac{4\pi}{\sigma}\right)^8 e^{-\frac{2}{\sigma}
\sqrt{b^2_1b_2^2-(b_1b_2)^2}-\frac{2}{\sigma}
\sqrt{b^2_3b_4^2-(b_3b_4)^2}}\equiv (2\pi)^4\delta (\Sigma p^{(i)})
I_4^{(0)}.\label{x50} \ee

The final expression for the 4-point function $\bar G_4(p_1,p_2, p_3, p_4)$
(where the factor $(2\pi)^4 \delta (\sum p_i)$ is separated out)

\be \bar G_4(p_i)=tr \prod^4_{i=1}\int d^4q_i \Gamma_i (m_i-i\hat
q_i)\int^\infty_0 e^{-s_i(m^2_1+q^2_1)} ds_i I_4^{(0)}(a,b) \exp (-
\mathcal{Q}). \label{x36} \ee The factor $\exp(-\mathcal{Q})$ in (\ref{x36}) is
  equal \be
\exp(-\mathcal{Q})=\exp\left[-\frac{g^2C_2}{8\pi^2}\sum_{i\leq j} I_{i,j} (s_i,
s_j, q_i, q_j)\right].\label{x37} \ee The vertex parts $I_{i,i+1}$ in
(\ref{x37}) may contain double logarithmic parts under the same conditions as
for the 3-point function. One can  see, that also in the 4-point case one
arrives at the asymptotics at large $b^2_i\gg \sigma$ \be J^{(0)}_4 (a,b) \to
\prod_{i=1,2,4} \delta^{(4)} (b_i).\label{x38}\ee

In this case the four $d^4q_i$ integrations in (\ref{x36}) reduce to a single
one, as it should be for the standard Feynman diagram without confinement.



\begin{thebibliography}{99}
%

\bibitem{1} R.P.Feynman, Photon-Hadron Interactions, W.A.Benjamin Inc. Reading MA, 1972.

\bibitem{2} B.L.Ioffe, V.A.Khose, and L.N.Lipatov, Deep Inelastic
Processes, North-Holland, 1984.
\bibitem{2a}  B.L.Ioffe, V.S.Fadin and L.Lipatov,
Quantum Chromodynamics, Cambridge University Press, Cambridge, U.K., (2010).

\bibitem{3} F.J.Yndurain, The Theory of Quark and Gluon Interactions, 4th edition,  Springer, 2006.

\bibitem{4} M.E.Peskin and D.V.Schroeder, ``Quantum Field
Theory'', Addison-Wesley Publishing Company, Reading 1995.

\bibitem{5} K.A.Olive et al., (Particle Data Group), Chin. Phys. {\bf C 38}, N9 (2014),
p.122.

\bibitem{5a}  S.Bethke, G.Dissertori and G.P.Salam, Quantum
Chromodynamics, p.296.

\bibitem{5b}  B.Foster, A.D.Martin and M.G.Vincter,
Structure functions.

\bibitem{6}  V.N.Gribov and
L.N.Lipatov, Sov. J. Nucl. Phys. {\bf 15}, 438 (1972).
\bibitem{6a}
Yu.M.Dokshitzer,  Sov. Phys.  JETP {\bf 46}, 641 (1977).
\bibitem{6b}  G.Altarelli and  G.
Parisi, Nucl. Phys. {\bf B 126}, 298 (1977).

\bibitem{7} G.Sterman, arXiv:1412.5698 [hep-ph].
\bibitem{7a}  A.V.Efremov and A.V.Radyushkin, Mod. Phys.
Lett.{\bf A 24}, 2803 (2009); arXiv: 0911.1195 [hep-ph].





\bibitem{8}

  H.G.Dosch, Phys. Lett. B {\bf 190}, 177 (1987).
  \bibitem{8a}
  Yu.A.Simonov, Nucl.  Phys.  B {\bf 307}, 512 (1988).
  \bibitem{8b}
  H.G.Dosch and Yu.A.Simonov, Phys. Lett. B {\bf 205}, 339 (1988).
  \bibitem{8c}
 A.Di Giacomo, H.G.Dosch, V.I.Shevchenko, Yu.A.Simonov, Phys. Rept. {\bf 372}, 319
 (2002).
 \bibitem{8d}
 Yu.A.Simonov, Phys. At. Nucl. {\bf 67}  846 (2004).
 \bibitem{8e}
Yu.A.Simonov,  Phys. At. Nucl. {\bf 67}  1027 (2004).
 \bibitem{8f}
 Yu.A.Simonov and J.A.Tjon,  Ann. Phys.  (N.Y.)  {\bf 300}, 54 (2002).

\bibitem{9}Yu.A.Simonov,   Phys.  Rev.    {\bf D 88},  025028 (2013).

\bibitem{9a} Yu.A.Simonov, Phys.  Rev.   {\bf D 90}, 013013  (2014).
\bibitem{10}Yu.A.Simonov,   Phys.  Rev.    {\bf D 88}, 053004 (2013).


\bibitem{11}Yu.A.Simonov, Phys. Rev. {\bf D 91}, 065001 (2015),  arXiv: 1409.4964 [hep-ph].

\bibitem{12}Yu.A.Simonov,  arXiv: 1411.7223, v.4 [hep-ph].


\bibitem{13} R.L.Jaffe and A.Manohar, Nucl. Phys. {\bf B 337}, 509 (1990).
\bibitem{13a}
X.Ji, Phys. Rev. Lett. {\bf 78}, 610 (1997),  hep-ph/9603249.
\bibitem{13b}  X.Ji, J.-H.Zhang
and Y.Zhao,  arXiv: 1409.6329.

\bibitem{13x} S.J.Brodsky, Nucl. Phys. Proc. Suppl. {\bf 90}, 3 (2000).



\bibitem{13xx} S.J.Brodsky, H.C.Pauli and S.S.Pinsky, Phys. Rep. {\bf 301}, 299 (1998).

\bibitem{14} Yu.A. Simonov,  Phys. At. Nucl. {\bf 67}, 553
(2004); hep-ph/0306310.



\bibitem{15}
Yu.A. Simonov, Phys. At. Nucl. {\bf 64}, 1876 (2001); hep-ph/0110033.
\bibitem{16}
R. Jaffe and K. Johnson, Phys. Lett.  {\bf B 60},  201  (1976).\bibitem{16a}
M.~S. Chanowitz and S.~R. Sharpe, Nucl. Phys.  {\bf B 222},  211
(1983).\bibitem{16b} T. Barnes, F. Close, and F. de~Viron, Nucl. Phys.  {\bf B
224},  241  (1983).\bibitem{16c} N. Isgur and J.~E. Paton, Phys. Rev.  {\bf D
31},  2910  (1985).

\bibitem{17}Yu.A. Simonov in: Proceeding of the Workshop on Physics and Detectors for
DA$\Phi$NE,  Frascati, 1991.

\bibitem{17a} Yu.A. Simonov, Nucl. Phys. B (Proc. Suppl.) {\bf
23 B}, 283 (1991),

\bibitem{17b} Yu.A. Simonov in: Hadron-93 ed. T. Bressani, A. Felicielo,
G. Preparata, P.G. Ratcliffe, Nuovo Cim. {\bf 107 A}, 2629 (1994).

\bibitem{17c} Yu.S.
Kalashnikova, Yu.B. Yufryakov, Phys. Lett.{\bf B  359}, 175 (1995).

\bibitem{17d} Yu.
Yufryakov, hep-ph/9510358.

\bibitem{17e} Yu.S.Kalashnikova and D.S.Kuzmenko, Phys. At. Nucl.
{\bf 67}, 538 (2004); hep-ph/0302070.

\bibitem{17f}
 Yu.S.Kalashnikova and A.V.Nefediev,  Phys.Rev.  {\bf D 77}, 0540025 (2008).

\bibitem{17*}
S. Ishida, H. Sawazaki, M. Oda, and K. Yamada, Phys. Rev.  {\bf D 47},  179
  (1993).

  \bibitem{17*a}
  T. Barnes, F. Close, and E. Swanson, Phys. Rev. {\bf D 52},  5242  (1995).

  \bibitem{17*b}
I.~J. General, S.~R. Cotanch, and F.~J. Llanes-Estrada, Eur.Phys.J. {\bf C 51},
  347  (2007).

  \bibitem{17*c}
I. Balitsky, D. Diakonov, and A. Yung, Z. Phys. {\bf C 33},  265  (1986).

\bibitem{17*d}
J. Latorre, P. Pascual, and S. Narison, Z. Phys. {\bf C 34},  347  (1987).

\bibitem{17e}
K.~G. Chetyrkin and S. Narison, Phys. Lett. {\bf B 485},  145  (2000).

\bibitem{17f}
S. Narison, Phys. Lett.{\bf  B 675},  319  (2009).

\bibitem{17*g}
 V.Mathieu, Phys. Rev. {\bf D 80}, 014016 (2009).

 \bibitem{17*h}L.S.Kisslinger, Phys. Rev.
 {\bf D 79}, 114026 (2009).
 \bibitem{17*i}
 C.Semay, F.Buisseret and D.Silvester-Brac, Phys. Rev. {\bf D 79}, 094020
 (2009).


\bibitem{18} P.Guo et al., Phys. Rev. {\bf D 77}, 056005 (2008).
\bibitem{18a} J.J.Dudek,
Phys. Rev. {\bf D 84}, 074023 (2011).

\bibitem{19}B.Ketzer, arXiv: 1208.5125.






\bibitem{20} S.Godfry and N.Isgur, Phys. Rev. {\bf D 32 }, 189 (1985).

\bibitem{21} F.Bloch and A.Nordsick, Phys. Rev. {\bf 52}, 54 (1937).

\bibitem{21a}
S. Weinberg, Phys. Rev. {\bf 140},  B 516 (1965).

\bibitem{21b} D.Yennie, S.Frautschi and
H.Suura,  Ann. Phys. {\bf 13}, 379 (1961).




\bibitem{22}  A.Le Yaouanc, L.Oliver, O.P\'{e}ne etal., Z. Phys. {\bf C 28},
309 (1985).

\bibitem{22a} F.Iddir, S.Safir and O.P\'{e}ne, Phys. Lett.  {\bf B 433}, 125
(1998).


\bibitem{23} A.Yu.Dubin, A.B.Kaidalov and Yu.A.Simonov, Phys. At. Nucl., {\bf
58}, 300 (1995), hep-ph/9408212.

\bibitem{23a} V.P.Morgunov, V.I.Shevchenko and Yu.A.Simonov,
Phys. lett. {\bf B 416}, 433 (1998).

\bibitem{24} CMS Collaboration, JHEP 1009 (2010) 091 [arXiv: 1009.4122
[hep-ex].
\bibitem{24a}   Wei Li, Mod. Phys. Lett. {\bf A27}, 1230018 (2012),
arXiv: 1206.0148 [nucl. ex].
 \bibitem{25a} M.Yu. Azarkin, I.M.Dremin and A.V.Leonidov, Mod. Phys. Lett.
 {\bf A 26},  963 (2011).
\bibitem{25b} M.Yu. Azarkin, I.M.Dremin and M.Strikman,  Phys. Lett.
 {\bf  B 735},  244 (2014).
 \bibitem{25c} L.Frankfurt , M.Strikman and C.Weiss,   Phys. Rev.
 {\bf D 83},  054012 (2011).


\bibitem{25} F.Gelis, E.Iancu, J.Jalilian-Marian, R.Venugopalan,  Ann. Rev. Nucl. Part. Sci. {\bf 60},
463 (2010); arXiv:1002.0333 [hep-ph].

\bibitem{26} Bo Andersson, S.Mohanty and F.S\"{o}derberg, arXiv:
hep-ph/0212122.

\bibitem{27} Yu.A.Simonov and A.I.Veselov, Phys. Rev. {\bf D 79}, 034024
(2009).

\bibitem{27a} I.V.Danilkin, V.D.Orlovsky and  Yu.A.Simonov, Phys. Rev. {\bf  D 85},
034012 (2012).

\bibitem{27b} A.M.Badalian, V.D.Orlovsky and  Yu.A.Simonov, Phys. At. Nucl.
{\bf 76}, 955 (2013).



\end{thebibliography}
\end{document}